\documentclass[12pt]{article}

\usepackage{amsmath}
\usepackage{amssymb}
\usepackage{amscd}

\newcommand{\F}{\mathbb F}
\def\Mat{{\mathrm{Mat}}}
\def\id{{\mathrm{id}}}

\begin{document}
\begin{center}
\large\bf
Cryptanalysis of some protocols using matrices over group rings\\
\end{center}
\begin{center}
\small Mohammad Eftekhari\\

\end{center}

\begin{center}
{\large\bf
Abstract}
\end{center}
{\it We address a cryptanalysis of two protocols based on the supposed difficulty of discrete logarithm problem on (semi) groups of
matrices over a group ring. We can find the secret key and break entirely the protocols.}\\


94A60, 16Gxx\\

{\bf\large 1.Introduction} \\

The Diffie-Hellman  key agreement  protocol is the first published practical solution to the key distribution problem, allowing two parties that have never met to exchange a secret key over an open channel. It uses the cyclic group $\F_q^*$, where $\F_q$ is the finite field with 
$q$ elements. The security of this protocol is based on the difficulty of computing discrete logarithms (DL) in the group $\F_q^*$.
There are several algorithms   for computing discrete logarithms, some of them are subexponential  when  applied
to $\F_q^*$.\\

 It is  important to search for easily implementable groups, for which the DL problem is hard and there is
no known subexponential time algorithm for computing DL. The group of points over $\F_q$ of an elliptic curve is such a group.\\
In [8], the group of invertible matrices with coefficients in a finite field was considered for such a key exchange. In [6], using the
Jordan form it was shown that the discrete logarithm problem on such matrices can be reduced to the same problem over
some small extensions of the finite base field.\\
In [4], the authors consider the semigroup of matrices ( 3-by-3 matrices) over the group ring $\F_7[S_5]$, where $S_5$ is the group
of permutation of $\{1,2,3,4,5\}$. The security of this protocol is based on the supposed difficulty of the discrete logarithm
problem in the (semi)  group  of matrices with coefficients in  $\F_7[S_5]$.\\
Moreover in [5], the authors propose the same semigroup as a platform for the Cramer-Schoup cryptosystem which is a generalization of ElGamal's protocol. Here the security is based on the supposed difficulty of the discrete logarithm problem
in the group of invertible 3-by-3 matrices with coefficients in  $\F_7[S_5]$.\\
In [1], [2] and [7] a cryptanalysis of [4] is proposed. Their methods are somehow different. In [1], the problem of discrete logarithm in a semigroup is reduced to the same problem in a subgroup of the same
semigroup. In [2] one uses a slight modification of Shor's quantum algorithm to find the period of a singular matrix (there is
no notion of order for such a matrix) and therby solving the discrete logarithm problem in semigroups. In [7], $\Mat_3(\F_7[S_5])$ 
is embeded in $\Mat_{360}(\F_7)$ and then one uses the same procedure as in [6] (adapted to singular matrices). The conclusion of all three papers above is that using a quantum computer one can break the key exchange protocol of [4].
\\
In contrast to the above analysis we
use the irreducible  representations of the group $S_5$; then using the fact that the algebra  $\F_7[S_5]$ is semi-simple, we
give an isomorphism between this algebra and an algebra of block matrices with coefficients in $\F_7$. Then we use this isomorphism to 
give an isomorphism between   $\Mat_3(\F_7[S_5])$, and still another algebra of block matrices over $\F_7$.
To do so, we combine the same blocks of the first isomorphism.\\
This way we reduce the discrete logarithm problem over  $\Mat_3(\F_7[S_5])$, to the same problem
over block matrices with coefficients in $\F_7$. The maximum size of a block is 18, reducing dramatically the computations. Now we can apply the same procedure (eventually modified for singular matrices) as in [4], to each block and resolve
the problem of discrete logarithm  entirely (using actual computers) and find  the secret key. So the conclusion is that the platform proposed in [4] and [5] are simply insecure.\\
The rest of this paper is organized as follows. Section 1, will be devoted to the irreducible representations of $S_5$.
In section 2, we explain the isomorphism between matrices with coefficients in  $\F_7[S_5]$, and block matrices with
coefficients in $\F_7$, and show that the protocols proposed in [4] and [5] can be broken. Finally we conclude with some remarks in section 3.\\

{\bf\large 2. Irreducible representations of $S_5$}\\
For our purpose, it will be easier to use the following presentation of $S_5$. We note $W:=(1 2)$ and 
$Z:=(1 2 3 4 5)$.
The group $S_5$ is defined by generators $W,Z$ and relations $T$,  where $T$ is the following set of relations:
$$\displaylines{
 W^2=\id\cr
 Z^5=\id\cr
 (ZW)^4=\id\cr
 WZ^{-1}WZW=Z^{-1}WZWZ^{-1}WZ\cr
 [W,Z^{-2}WZ^2]=\id\cr
 [W, Z^{-3}WZ^3]=\id\cr
}$$

The group $S_5$ has two distinct representations of dimension one (namely the trivial one and the signature), two non isomorphic irreducible representations
of dimension four, two non isomorphic irreducible representations of dimension five, and one irreducible representation of dimension six. We give the images of the generators $Z$ and $W$ by these representations , and one can verify the
relations $T$, for the images, thereby proving that one defines morphisms from $S_5$ to matrix groups. One can compare the
trace of these morphisms with the character table of $S_5$ to be sure we obtain all the irreducible representations of
$S_5$.\\
To construct these representations one can follow the general description of [3], using Young polytabloids, to construct
the Specht modules which give the irreducible representation of $S_5$.\\

 $W=(1 2)\longmapsto A_1\oplus A'_1\oplus A_4\oplus A'_4\oplus A_5\oplus A'_5\oplus A_6$ where\\

$A_1=1$; $A'_1=-1$;$A_4=\left(\begin{array}{cccc}
-1&0&0&-1\\
0&-1&0&1\\
0&0&-1&-1\\
0&0&0&1
\end{array}\right )$;
$A'_4=\left(\begin{array}{cccc}
1&0&0&1\\
0&1&0&-1\\
0&0&1&1\\
0&0&0&-1
\end{array}\right )$\\
$A_5=\left(\begin{array}{ccccc}
-1&0&1&0&-1\\
0&-1&-1&0&0\\
0&0&1&0&0\\
0&0&0&-1&-1\\
0&0&0&0&1
\end{array}\right )$;
$A'_5=\left(\begin{array}{ccccc}
1&0&-1&0&1\\
0&1&1&0&0\\
0&0&-1&0&0\\
0&0&0&1&1\\
0&0&0&0&-1
\end{array}\right )$\\
$A_6=\left(\begin{array}{cccccc}
-1&0&1&0&1&0\\
0&-1&-1&0&0&1\\
0&0&1&0&0&0\\
0&0&0&-1&-1&-1\\
0&0&0&0&1&0\\
0&0&0&0&0&1
\end{array}\right )$\\

$Z=(1 2 3 4 5)\longmapsto  B_1\oplus B'_1\oplus B_4\oplus B'_4\oplus B_5\oplus B'_5\oplus B_6$ where\\

$B_1=1$; $B'_1=1$;
$B_4=\left(\begin{array}{cccc}
0&0&0&1\\
-1&0&0&-1\\
0&-1&0&1\\
0&0&-1&-1
\end{array}\right )$;
$B'_4=\left(\begin{array}{cccc}
0&0&0&1\\
-1&0&0&-1\\
0&-1&0&1\\
0&0&-1&-1
\end{array}\right )$\\
$B_5=\left(\begin{array}{ccccc}
0&0&-1&-1&-1\\
0&0&0&1&0\\
0&0&0&-1&-1\\
1&0&-1&-1&0\\
0&1&1&1&1
\end{array}\right )$;
$B'_5=\left(\begin{array}{ccccc}
0&0&-1&-1&-1\\
0&0&0&1&0\\
0&0&0&-1&-1\\
1&0&-1&-1&0\\
0&1&1&1&1
\end{array}\right )$\\
$B_6=\left(\begin{array}{cccccc}
0&0&1&0&0&0\\
0&0&0&0&1&0\\
0&0&0&0&0&1\\
1&0&-1&0&-1&0\\
0&1&1&0&0&-1\\
0&0&0&1&1&1
\end{array}\right )$\\\\
\vskip  5mm

{\bf\large 3. Cryptanalysis of protocols}\\

In [4]  the authors propose the Diffie-Hellman key exchange using 3-by-3 matrices over $\F_7[S_5]$. So Alice and Bob,
take a public matrix $M\in \Mat_3(\F_7[S_5])$ which may be non-invertible. Alice chooses a secret integer $n$, computes
$M^n$ and sends it to Bob. Bob chooses a secret integer $n'$, computes $M^{n'}$ and sends it to Alice. Every party can now
compute the common key $M^{nn'}$.\\
In [5], they use the same platform for the Cramer-Schoup cryptosystem which we do not recall. We underline only that there
is a public key $M$ as above, and during the protocol among other data sent, there is $M^n$ where $n$ is the secret key.
So if we are able to give a solution for the discrete logarithm problem in the case of $M\in \Mat_3(\F_7[S_5])$, in both cases the platform proposed is not secure. That is what we are going to explain.\\
As $7$ does not divide $|S_5|=120$, the algebra $\F_7[S_5]$ is semi-simple and Maschke's theorem asserts that this
algebra is isomorphic to a direct sum of matrix algebras (over $\F_7$), in other words it is isomorphic to an algebra of
block matrices over $\F_7$. Let us denote by  $f$  this isomorphism. To be of any use for our purpose, we have to make precise 
this isomorphism explicitly. The $\F_7$-linear extension (to $\F_7[S_5]$) of the morphism of $S_5$ using the irreducible representations of $S_5$ given on generators $W=(1 2), Z=(1 2 3 4 5)$ in section 2, gives the isomorphism $f$ between $\F_7[S_5]$ and its image. So for any element $x=\sum_{i=1}^{120}a_ix_i\in \F_7[S_5]$ , $a_i\in \F_7$ and $x_i\in S_5$ we can compute its image as a direct sum of matrices with coefficients in $\F_7$.\\
Up to now we have represented a matrix $M\in \Mat_3(\F_7[S_5])$ as a matrix with coefficients in $\F_7$ by replacing each coefficient
 $M_{ij}$ of $M$ by $f(M_{ij})$. For example $M_{11}$ is replaced by 
$A=\left(\begin{array}{ccccccc}
a_1&&&&&&\\
&a'_1&&&&&\\
&&a_4&&&&\\
&&&a'_4&&&\\
&&&&a_5&&\\
&&&&&a'_5&\\
&&&&&&a_6\\

\end{array}\right )$ where $a_i,a'_i$ are block matrices with coefficients in $\F_7$ and the indices denote the size of the block.

Let us denote  by $A,B,C,E,F,G,H,I,J$ the block matrices corresponding to $M_{11},M_{12},M_{13},M_{21},M_{22},M_{23},M_{31},M_{32},M_{33}$. Then $B$ is a block matrix which we represent the same way as $A$ by denoting $b_1,b'_1,b_4,b'_4,...$ its blocks. We use the same notations for $C,D,...$. It is an easy computation to prove that there is
a natural isomorphism between matrices\\
$\left(\begin{array}{ccc}
A&B&C\\
D&E&F\\
H&I&J
\end{array}\right )$ and the block matrix whose first block is obtained by composing (side by side) the first blocks of $A,B,C,D,...$
,namely 
$\left(\begin{array}{ccc}
a_1&b_1&c_1\\
d_1&e_1&f_1\\
h_1&i_1&j_1
\end{array}\right )$, which gives a $3\times 3$ matrix over $\F_7$.\\
 The second block is obtained by composing the second blocks of $A,B,C,D,...$, namely
$\left(\begin{array}{ccc}
a'_1&b'_1&c'_1\\
d'_1&e'_1&f'_1\\
h'_1&i'_1&j'_1
\end{array}\right )$, and so on.\\
To resume, we represent the matrix $M\in \Mat_3(\F_7[S_5])$ by a block matrix in $\F_7$ whose blocks are of size
$3,3,12,12,15,15,18$. We represent also the matrix $M^n$ by a block matrix with the same size $3,3,12,12,15,15,18$ in
$\F_7$. Now we can apply the same technics as in [6], namely write the Jordan form of each block in some small extension base 
$\F_{\displaystyle 7^\alpha}$ and find the secret key $n$. Note that for singular blocks, we need a slight modification of the
procedure of [6], as proposed in [7].
\\
\vskip 5mm

{\bf\large 3. Conclusion}\\

We showed that using matrices with coefficients in $\F_7[S_5]$ as a platform for Diffie-Hellman key exchange is not secure.
One may wonder if replacing $\F_7$ by $\F_2,\F_3$ or $\F_5$ give something essentially different. In fact in these cases
the group algebra is not semi-simple anymore and Wederburn's theorem cannot be applied. But these new algebras are not far from being semi simple; in fact they differ from being semi simple by a nilpotent radical, and the quotient is semi simple and then the same
procedure as explained in section 2 can be applied. To resume we believe that no secure cryptographic protocol can be based upon
these algebras.\\
Furthermore replacing the group $S_5$ by some other finite group $G$, can be cryptanalyzed the same way using the irreducible representations of $G$.\\

{\bf\large 3. References}\\

[1] M. Banin, B. Tsaban,
A reduction of semigroup DLP to classic DLP,
arXiv eprint 1310.7903.\\

[2] A. Childs, G. Ivanyos,
Quantum computation of discrete logarithms in semigroups,
arXiv eprint 1310.6238.\\

[3] G.D. James, The representation theory of the symmetric groups, SLN 682, 1978. \\

[4] D. Kahrobaei, C. Koupparis, W. Shpilrain, Public key exchange using matrices over group rings, G.C.C. Vol. 5, Issue 1. 2013, p. 97-115.\\
\pagebreak

[5] D. Kahrobaei, C. Koupparis, W. Shpilrain, A CCA secure cryptosystem using matrices over group rings, Contemp. Math., Amer. Math. Soc. 633 (2015), p. 73-80. 

[6] A.J. Menezes , YI-H. Wu , The discrete logarithm problem in $GL_n(\F_q)$, ARS combinatorica. 1997. vol. 47. p. 23-32.\\

[7] A. Myasnikov, A. Ushakov, Quantum algorithm for discrete logarithm problem for matrices over finite group rings,
Journal of Symbolic Computation, to appear, available online.\\

[8]  R.Odonne, D. Varadharajan, P. Sanders, Public key distribution in matrix rings, Electronic letters, 20 (1984),p. 386-387.\\
\vskip 15mm

\noindent {\bf Author information}\\
\noindent Mohammad Eftekhari, 
LAMFA, CNRS UMR 7352, 
 Universit\'e  de Picardie-Jules Verne, 33 rue Saint-Leu 80039 Amiens France\\
E-mail: mohamed.eftekhari@u-picardie.fr\\

\end{document}